\documentclass[useAMS,usenatbib]{mn2e}

\usepackage{float} 
\usepackage{epsfig}
\usepackage{graphicx}
\usepackage[latin1]{inputenc}
\usepackage{latexsym}
\usepackage{multirow}
\newcommand{\langl}{\begin{picture}(4.5,7)
\put(1.1,2.5){\rotatebox{60}{\line(1,0){5.5}}}
\put(1.1,2.5){\rotatebox{300}{\line(1,0){5.5}}}
\end{picture}}
\newcommand{\rangl}{\begin{picture}(4.5,7)
\put(.9,2.5){\rotatebox{120}{\line(1,0){5.5}}}
\put(.9,2.5){\rotatebox{240}{\line(1,0){5.5}}}
\end{picture}}

\title[CL0217 observations with SRT]{Observations of the galaxy cluster CL 0217+70 and its surrounding region at 1.4 GHz with the Sardinia Radio Telescope}
\author[P. Marchegiani et al.]{P. Marchegiani$^{1}$\thanks{E-mail: paolo.marchegiani@inaf.it}, V. Vacca$^{1,2}$, F. Govoni$^{1}$, M. Murgia$^{1}$, F. Loi$^{1}$, and L. Feretti$^{3}$\\ \\ 
$^{1}$INAF-Osservatorio Astronomico di Cagliari, Via della Scienza 5, I-09047 Selargius (CA), Italy\\
$^{2}$Max Planck Institute for Astrophysics, Karl-Schwarzschildstr. 1, 85741 Garching, Germany\\
$^{3}$INAF-Istituto di Radioastronomia, via P. Gobetti 101, 40129 Bologna, Italy\\
}

\begin{document}

\date{Accepted 2024 November 21. Received 2024 October 31; in original form 2024 February 15}

\pagerange{\pageref{firstpage}--\pageref{lastpage}} \pubyear{2024}

\maketitle

\label{firstpage}

\begin{abstract}
We present the results of observations performed with the Sardinia Radio Telescope (SRT) at 1.3--1.8 GHz of the galaxy cluster CL 0217+70 and a $3^\circ \times 3^\circ$ region around it. We combine the SRT data with archival Very Large Array (VLA) data to obtain images having the VLA angular resolution, but sensitive up to largest scales. The SRT+VLA combination allows us to derive a cluster radio halo flux density higher by $\sim14\%$ compared to the VLA-only data, although consistent within $1\sigma$. We derive a spectral index map between 140 MHz and 1.4 GHz, finding an extended region with spectral index $\alpha\sim0.6$ on the external part of the south-eastern candidate relic, questioning the real nature of this relic. Moreover, we detect an extended emission outside the cluster in the south-eastern area, having an angular extension of $\sim50$ arcmin on the longer side, which would correspond to $\sim10$ Mpc at the cluster distance; the emissivity that this region would have if located at the cluster distance is in line with the one estimated in candidate filaments of the cosmic web; however, the peculiar orientation of this region, not pointed towards the cluster, and the low Galactic latitude of this cluster suggest that its origin can be due to a foreground emission originating in our Galaxy.
\end{abstract}



\section{Introduction}

The study of galaxy clusters in the radio band has shown the presence in a fraction of them of diffuse emissions, not directly related to single galaxies, with different location, morphological, and polarization properties; among them, radio halos are located in the central regions of the clusters, approximately roundish, and with a low level of polarization, while radio relics are located at the periphery of the clusters, elongated, and with a higher level of polarization (e.g. Feretti et al. 2012; van Weeren et al. 2019). These radio structures, being produced by synchrotron emission, indicate the presence on large scale of relativistic electrons and magnetic fields, which in turn require the presence of mechanisms that can produce and accelerate relativistic particles (e.g. Jaffe 1977; Blasi \& Colafrancesco 1999; Dolag \& En\ss lin 2000; Brunetti et al. 2001; see also Brunetti \& Jones 2014 for a review).

According to our present understanding of the formation of structures in the universe, galaxy clusters form at the intersection of over-dense filaments by merging of smaller structures and accretion of external matter; for this reason, they are not isolated structures, but often exhibit evidence of connections and interactions with other surrounding structures or the cosmic web. Recently, the study of these larger scale structures in the radio band has allowed to detect the presence of diffuse emissions on large scales in clusters in total intensity (Cuciti et al. 2022) and in polarization (Vacca et al. 2022), bridges observed between interacting clusters (Govoni et al. 2019; Botteon et al. 2020) and in stacked analysis (Vernstrom et al. 2021, 2023), and hints of emission in large scale filaments (Vacca et al. 2018), suggesting that magnetic fields and mechanisms of acceleration of relativistic particles are present also on large scales outside the virial radius of clusters.

The spatial distribution of the surface brightness of diffuse radio emissions in galaxy clusters is an important source of information regarding the properties of the non-thermal electrons (e.g. Govoni et al. 2001), of the acceleration mechanisms of the electrons (e.g. Brunetti et al. 2001), of the magnetic field (e.g. Murgia et al. 2004, 2009), and possibly of other components of the cluster as non-thermal protons (e.g. Marchegiani, Colafrancesco \& Perola 2007) and dark matter (e.g. Marchegiani \& Colafrancesco 2016), in the case the electrons are secondary products of hadronic interactions or dark matter annihilation respectively.

In the case of a nearby and large galaxy cluster like Coma, an angular resolution just sufficient to derive some useful indications about the radio surface brightness spatial distribution can be obtained using large single dish telescopes (see e.g. Deiss et al. 1997 and Thierbach et al. 2003, who studied the Coma cluster at 1.4, 2.675, and 4.85 GHz using the 100 m Effelsberg telescope; see also Murgia et al. 2024 for a recent study of the Coma cluster at 6.6 GHz with the Sardinia Radio Telescope). However, for more distant clusters an angular resolution of the order of 9--10 arcmin, as the one obtained at 1.4 GHz with a 100 m telescope, is not enough to appreciate with sufficient detail the internal structure of the diffuse radio emission and disentangle it from discrete embedded sources; therefore interferometric observations are necessary.

While interferometric observations can provide information until very small angular scales, their property of having a non-zero minimum baseline makes them not sensitive to structures larger than a certain scale. One possible way to obtain information on both small and large angular scales is to perform a combination of measures obtained with interferometric and single dish instruments (e.g. Loi et al. 2017; Vacca et al. 2018; Murgia et al. 2024). For this purpose, the size of the single dish instrument must be larger than the minimum baseline of the interferometer, so that there is a range of lengths, or equivalently of angular scales, where both the instruments are sensitive, allowing to adjust their flux scales so that they can overlap each other, reducing issues due to possible differences in calibration. The Sardinia Radio Telescope (SRT; Bolli et al. 2015; Prandoni et al. 2017), with its 64 meters of diameter, is a single dish instrument optimal for this purpose, allowing to obtain information on large scales about galaxy clusters and the surrounding regions, and to combine this information with higher resolution data coming from interferometers as the National Radio Astronomy Observatory (NRAO) Very Large Array (VLA).

In this paper we apply this idea to study the galaxy cluster CL 0217+70 (in the following CL0217) and the surrounding regions. This massive and hot cluster ($M\sim10^{15}~$M$_\odot$, $kT\sim 8$ keV) has first been identified in the Westerbork Northern Sky Survey (WENSS; Rengelink et al. 1997), and studied in detail by Brown, Duesterhoeft \& Rudnick (2011), who detected diffuse radio emissions in the form of a central radio halo, several relics or candidate relics, and other internal filaments observing with the VLA at the frequencies of 0.3 and 1.4 GHz. They also identified the cluster in the optical band as an overdensity in the Sloan Digital Sky Survey (SDSS), from where they estimated a photometric redshift of $z\sim0.065$, and in the X-ray in the ROSAT All Sky Survey (RASS; Voges et al. 1999). 

The values of the X-ray luminosity $L_X$ and radio power $P_R$ obtained for that value of the redshift had indicated that this cluster was an outlier in the $L_X/P_R$ correlation followed by most of the radio halo clusters (Feretti et al. 2012), 
being underluminous in the X-ray compared to what can be expected from its radio power.
However, a recent observation of the position of the iron lines in the cluster X-ray spectrum obtained with Chandra (Zhang et al. 2020) has allowed to redetermine the value of the redshift ($z\sim0.18$); using this value of the redshift, it is possible to 
derive values of $L_X$ and $P_R$ that are consistent with the correlation followed by other clusters.

Recently the cluster has been observed again in the radio band with the LOw Frequency ARray (LOFAR) at 141 MHz and with the VLA at 1.5 GHz by Hoang et al. (2021), who confirmed the presence of the central radio halo and two peripheral relics, with the north-western one having an elongated shape, and the south-eastern one having a more complicated hook-like shape, possibly divided in different parts, which, if belonging to an unique structure, would have a total length of 3.5 Mpc. These sources are located at a distance of about 3 Mpc from the cluster centre. The cluster has also been studied at 1.5 GHz with the VLA by Stuardi et al. (2022), with a particular focus on the polarization in the relic regions, finding that the north-western relic shows a maximum fractional polarization of $23\%$, while in the south-eastern relic the only region where a polarization is detected corresponds to the lobes of a radio galaxy, whereas in the other parts of this structure only upper limits in polarization have been found.

At the moment therefore this cluster has never been studied with a single dish instrument; in this paper we present the results of observations of this cluster performed at 1.4 GHz with the SRT, in order to explore not only the region of the cluster, but also the surrounding regions, with the aim of obtaining information on the presence of relativistic particles and magnetic fields on large scale. Since the angular resolution of SRT at 1.4 GHz, approximately 14 arcmin, is not sufficient to resolve the structures on smaller scale, we also combine these results with existing VLA observations, in order to have an angular resolution of the order of 1 arcmin, making it possible to resolve also small structures.

The plan of the paper is the following: in Section 2 we present the details of the observations used in this paper and the process of data reduction followed for obtaining the resulting radio maps with SRT and VLA separately; we also compare these maps with available X-ray maps. In Section 3 we present the procedure and the results of the combination of SRT data with data coming from the NRAO VLA Sky Survey (NVSS) and from pointed VLA observations. In Section 4 we inspect the combined SRT+VLA image, studying the properties of the radio halo of the cluster and other extended structures external to the cluster, in particular a wide region located south-east of the cluster, comparing these results with the information obtainable from the VLA-only image. 
Finally, we summarize our conclusions in Section 5.

We adopt a cosmological model with $H_0=70$ km s$^{-1}$ Mpc$^{-1}$, $\Omega_M=0.3$, and $\Omega_\Lambda=0.7$; with this model at the redshift of the cluster ($z=0.18$) 1 arcmin corresponds to $\sim182$ kpc, while the luminosity distance of the cluster is $D_L\sim872$ Mpc.

\section{Radio observations}

Details of observations are summarized in Tables \ref{tab.data} and \ref{tab.data.vla}. In the following we describe the procedures we have followed for collecting and reducing the data.

\begin{table*}{}
\caption{Details of SRT observations}
\begin{center}
\begin{tabular}{|*{8}{c|}}
\hline 
Frequency & Number of    & Time on  & Center of pointing &  OTF scans & FOV & Program & Observation date \\
 & channels &  source & (J2000) & & & &\\
\hline 
1.3--1.8 GHz & 16384  & 5h & 02:17:00 +70:36:15 & 3RA$\times$3DEC & 3$^\circ\times3^\circ$ & S0001 & 10 July 2016 \\
\hline
 \end{tabular}
 \end{center} 
 \label{tab.data}
 \end{table*}
 
 \begin{table*}{}
\caption{Details of VLA observations}
\begin{center}
\begin{tabular}{|*{8}{c|}}
\hline 
Frequency  & Configuration  & Time on & Center of pointing & Original & Original noise & Program & Observation date \\
(MHz) & & source & (J2000) & beam & (mJy/beam) & & \\
\hline 
1372.6--1410.1  & \multirow{2}{*}{D} &  \multirow{2}{*}{3.5h} & \multirow{2}{*}{02:17:00 +70:36:36} & \multirow{2}{*}{48''$\times$42''} & \multirow{2}{*}{0.1} & \multirow{2}{*}{AD509} &  \multirow{2}{*}{4,8,12 December 2005}\\
1452.4--1489.9  & & & & & & & \\ \\
\multirow{2}{*}{1008.0--2032.0}  & \multirow{2}{*}{D} & 55 min & 02:14:31 +70:41:04 & 60''$\times$39'' & 0.1 & \multirow{2}{*}{17A-083} & \multirow{2}{*}{20--21 March 2017}\\
 &  & 47 min & 02:18:50 +70:27:36 & 60''$\times$39'' & 0.09 & & \\
\hline
 \end{tabular}
 \end{center} 
 \label{tab.data.vla}
 \end{table*}

\subsection{SRT observations}

The radio observations presented in this paper, whose properties are summarized in Table \ref{tab.data}, were performed in the context of the SRT Multi-frequency Observations of Galaxy Clusters program (S0001; see Govoni et al. 2017 for a description of the project).
Data have been acquired on 10 July 2016 using the SARDARA backend at SRT (Melis et al. 2018), by performing on-the-fly scans with angular extension of 3 degrees centered on the cluster (three along RA and three along DEC) for a total observation time of about 5 hours.  
The frequency band of the data is 1.3--1.8 GHz, and they are recorded inside a wider band between 800 and 2300 GHz divided in 16384 channels ($\approx 90$ kHz each channel), with data occupying the central third. Data reduction has been performed using the SCUBE software (Murgia et al. 2016). The flux density scale used in the calibration process is the one from Perley \& Butler (2017). 

Data have been first flagged for RFI by using known information about bad channels as previously identified at the SRT site; after this operation, an auto-flag procedure searching for channels with anomalous emission with respect to nearby channels and other scans has been applied to the calibrator data, allowing to flag about 10\% of data. 

The source 3C147 has been used as primary flux calibrator; the baseline level for this source has been found and then removed by performing a linear fit to the background, and the bandpass correction has been calculated and applied.
The calibrator scans have then been imaged and modeled using a bi-dimensional Gaussian in order to derive the value of the source peak in units of counts.
This value has allowed to obtain the value of the conversion factor between counts and mJy/beam units, which is necessary to perform the data calibration in physical units. 

The removal of the baseline in the cluster data has been performed in two steps: in the first step we obtained from the NVSS (Condon et al. 1998) a mosaic image with size of 4 degrees per side, centered on the center of the SRT observed field and convolved with the SRT beam (14 arcmin); 
we used this image to model the radio sources in the field and the cold sky.
After this first baseline removal, the automatic flagging procedure has been applied to the cluster data to search for channels dominated by RFI,
allowing to flag $\sim8.9\%$ of data.
Subsequently, clusters data have been gridded for each scan separately, and finally different scans have been averaged in order to obtain the final image by estimating the correct weight to give to each scan depending on its goodness based on the local noise,
following the Stationary Wavelet Transform (SWT) procedure as described in Murgia et al. (2016).

\begin{figure*}
\centering
\begin{tabular}{c}
\includegraphics[width=0.53\textwidth, trim={1.5cm 9.5cm 0cm 0cm}, clip]{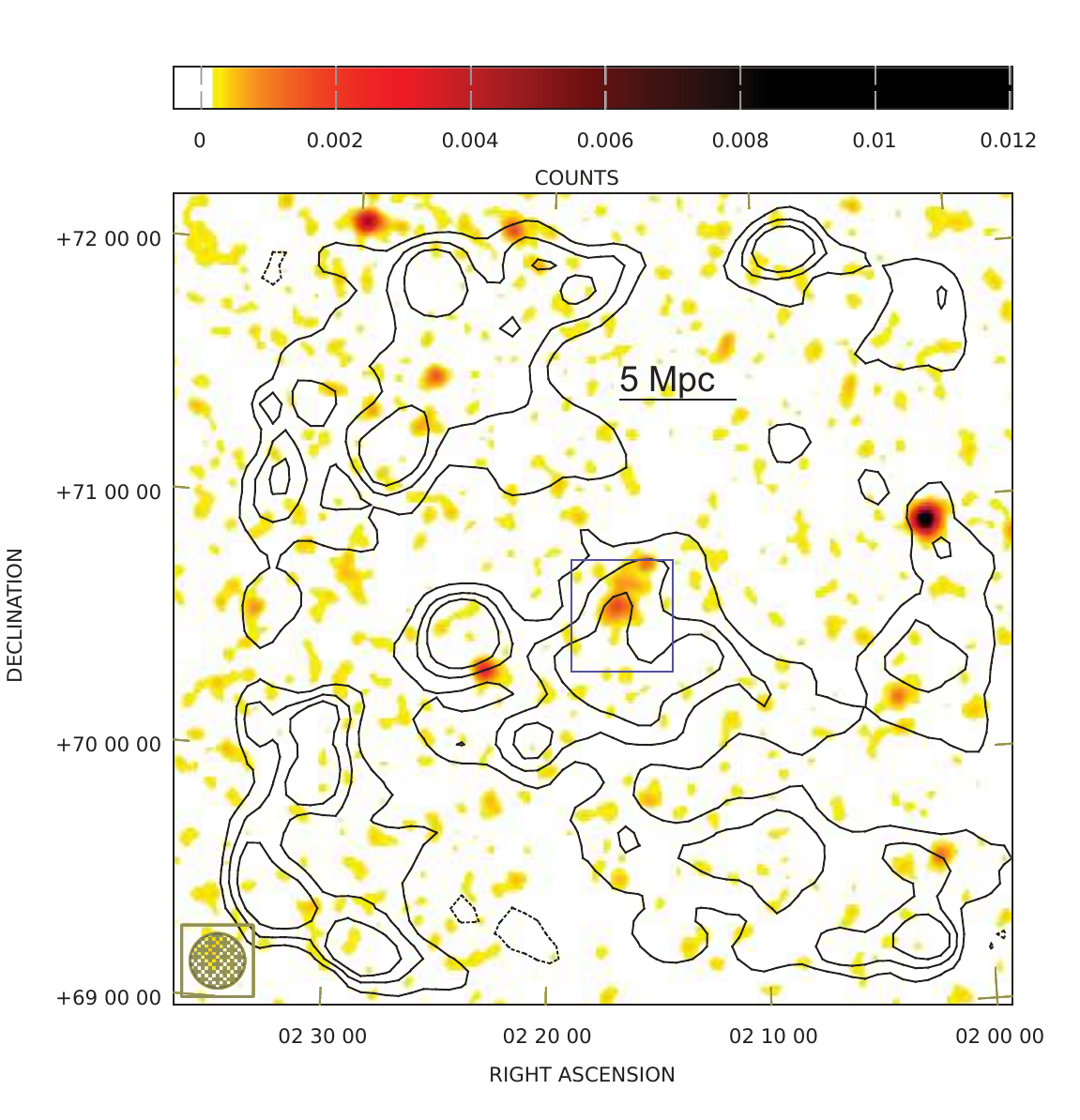}
\includegraphics[width=0.47\textwidth, trim={1.5cm 9.3cm 2cm 0cm}, clip]{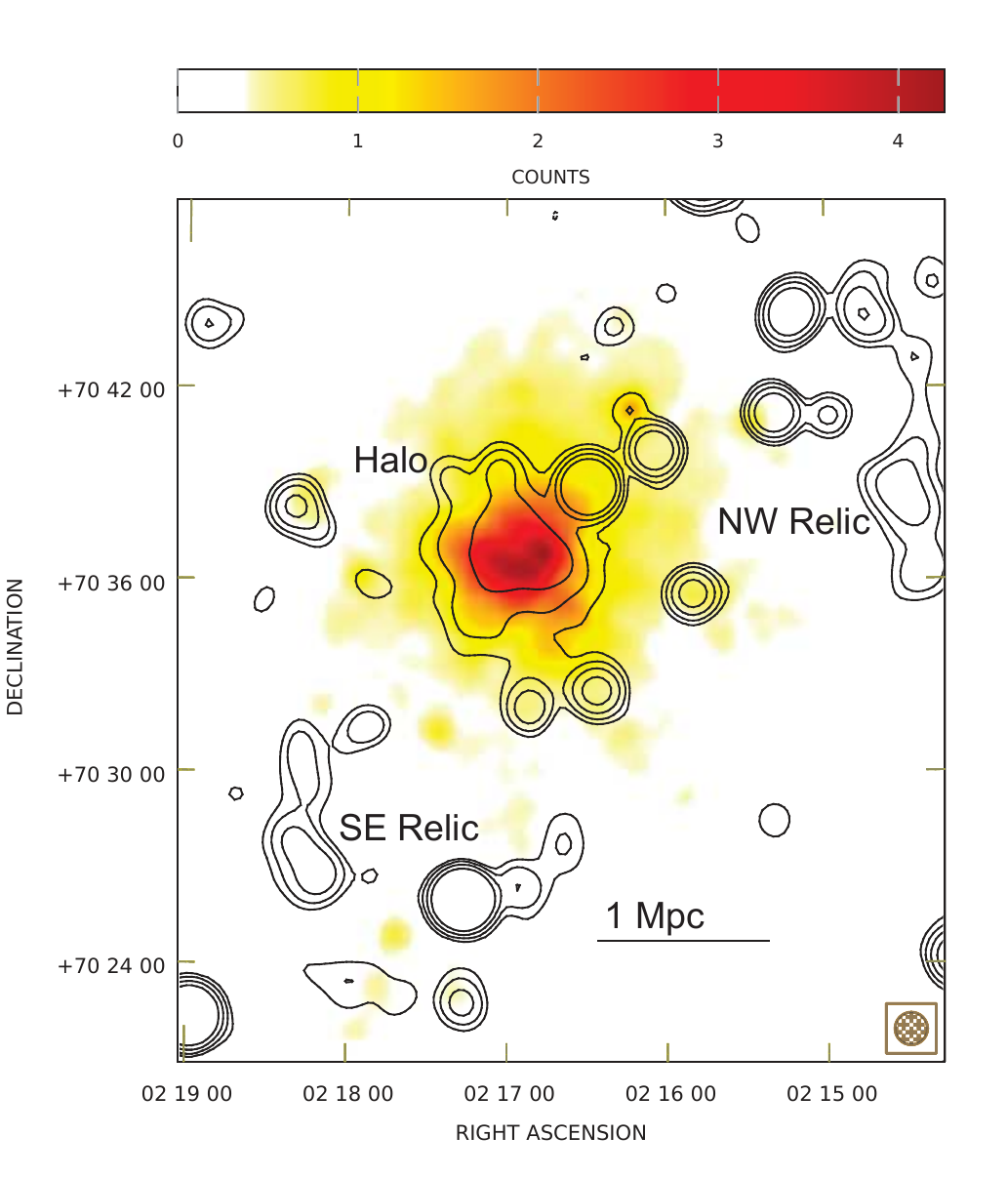}
\end{tabular}
\caption{\textit{Left panel:} 0.1--2.4 keV RASS image of the region around the cluster CL0217, with SRT contours overlapped. The horizontal bar indicates a length of 5 Mpc at the cluster distance. SRT contours are traced at levels of 0.045, 0.09, and 0.135 Jy/beam (positive contours, solid) and -0.045 Jy/beam (negative contours, dotted), i.e. at multiple values of $3\sigma$, with $\sigma=15$ mJy/beam. The SRT beam (FWHM 784 arcsec) is plotted in the bottom-left corner. The RASS image is smoothed with a 5-pixel (corresponding to 225 arcsec) Gaussian beam. The box contains the cluster region, which is zoomed in the right panel. \textit{Right panel:} 0.5--7 keV Chandra image of the cluster region with VLA contours overlapped. Contour levels start at $3\sigma$, with $\sigma=0.15$ mJy/beam, and scale with a factor of two. The horizontal bar indicates a length of 1 Mpc at the cluster distance. The VLA beam (FWHM 61 arcsec) is plotted in the bottom-right corner. The Chandra image is smoothed with a 5-pixel (corresponding to 20 arcsec) Gaussian beam.}
\label{fig.radio_Xray}
\end{figure*}

In the second step, the image obtained as a result of the first step has been used as a model image to better identify the regions where to compute the baseline level. After this more accurate estimation of the baseline we repeated the same procedures as in the first step. The final image, obtained by integrating along all the frequency band covered by data, is shown as contours in Fig.\,\ref{fig.radio_Xray}, left panel.
We have estimated the size of the SRT beam by performing a Gaussian fitting to the calibrator image obtained in four cross scans and averaging the results, obtaining a FWHM value of 784 arcsec (i.e. $\sim13.1$ arcmin). We note that during the observations the cluster was at an elevation in the range $\sim50-60$ degrees, while the calibrator 3C147 was at an elevation of $\sim42-43$ degrees.
In this elevation range the SRT gain curve in the L-band is basically flat (see Bolli et al. 2015), therefore no corrections for this difference are necessary.

The estimated noise in the image is 15 mJy/beam; this value is smaller than the confusion noise that can be estimated using theoretical approximations at a frequency of 1.55 GHz and for a beam of 13.1 arcmin, which, for a spectral index of $-0.7$, is $\sigma_c\sim25$ mJy/beam using the analytical expression of Condon (2002), and $\sigma_c\sim40$ mJy/beam using the results of numerical simulations of Loi et al. (2019); since we did not perform any cleaning procedure that could have artificially lowered the noise, we estimate that the more reliable estimate of the noise is the one obtained from the image rather than from theoretical extrapolations. Therefore, the SRT map is confusion limited, but we note that the extended structures visible in Fig.\,\ref{fig.radio_Xray} have a surface brightness higher than the confusion noise (the first contour is traced at 45 mJy/beam).

As it is possible to see, data cover a wide area, approximately a square area with size of 3 degrees per side, where the cluster is located at the center of the image. As expected, SRT observations alone are not able to well resolve the diffuse emissions located inside the cluster as the halo and the relics. It is instead possible to see the presence of large scale emissions, whose nature deserves to be better examined. For this purpose it is important to perform the combination with interferometric data, in order to resolve the structures on small scales and estimate the contribution from discrete sources; it is also useful to compare the radio maps with X-ray ones, in order to investigate the possible presence of other cosmic structures located in the wide area surrounding this cluster.

For this purpose, we also show in left panel of Fig.\ref{fig.radio_Xray} an image from the RASS with the SRT contours overlapped. The cluster is visible at the center of the X-ray image as a slightly extended emission with an evident elongation in the north-west direction, where another X-ray peak is located, approximately at the position (02:15:30 +70:46:30); this second peak could indicate the presence of a sub-cluster that is interacting with the main one, but it can also be a point source (we note that Brown et al. 2011 searched for counterparts in optical and radio bands, finding no obvious counterparts). This system appears to be surrounded by an extended radio emission, which seems to extend well beyond the central region, in particular in the south-east direction. At larger distance from the center of the image, close to the north-east, south-east, and south-west corners, there are other regions with extended radio emissions not immediately associated with X-ray sources. 
Given the SRT resolution, it is not possible to say if they are produced just as the blend of discrete sources, or if there are also extended components embedded in these structures. Given their proximity to the edges of the images, and in particular to the corners, we also can not exclude they are artifacts. In the rest of the paper we do not focus on these regions, even because they are not covered by the VLA data, and the combination procedure would not be possible for them. 
We also verified that in these regions there are not structures that, in X-rays or other spectral bands, are identified as other possible galaxy clusters.

\subsection{VLA observations}

We use archival VLA data at L-band in
D configuration in the direction of the galaxy cluster CL0217
in the context of the observing programs AD509 and
17A-083; their properties are summarized in Table \ref{tab.data.vla}.

The observing program AD509 consists of one pointing covering the central region of the cluster
in the frequency ranges 1372.6--1410.1\,MHz and 1452.4--1489.9\,MHz (3 channels per subband).
The total duration of the observation is about 3.5\,h.
The data were reduced following standard procedures using the NRAO's Astronomical
Image Processing System (AIPS) package. The used calibrator is 3C286.

The observing program 17A-083 consists of two pointings respectively centered on the south-eastern and on the north-western relic
in the 1008.0--2032.0\,MHz band (16 subbands, 64\,MHz and 64 channels each).
The total observing time is about 55\,min for the west pointing and 47\,min for the east one.
The data were reduced using the \textsc{common astronomy software applications (casa)}, pipeline version 5.3.1.
The used calibrators are 3C138 and 3C147.

For both datasets, we adopted the flux density scale by Perley \& Butler (2017) and performed
self-calibration and imaging with AIPS. The final image has been obtained by selecting the same
frequency range in both datasets and convolving to a full width half maximum (FWHM) of 61 arcsec.
The noise of the image estimated after this procedure is 0.15\,mJy/beam; note that this value has been estimated in the primary beam corrected image and that, since this a mosaic image, it has not been possible to estimate the noise without the primary beam correction. The noise level is found to increase only close the edges of the image; in the following analysis, we do not consider these peripheral regions.

In Fig.\ref{fig.radio_Xray}, right panel, we show an image obtained from Chandra archival data (Observation ID: 16293) overlapped with VLA contours. This image covers a smaller area centered on the cluster region compared to the SRT one, shown with the box in the left panel of the same Figure; it shows the X-ray slightly elongated and irregular structure of the cluster, and the known diffuse structures, as the central halo, the north-western relic, and the south-eastern structures that can be part of another giant relic.

\section{Combination of single dish and interferometric images}

In this section, we perform the combination of the SRT measures with data coming from interferometric observations. 
Data coming from pointed VLA observations do not cover the whole SRT observed field, but are limited to a region immediately surrounding the cluster. For this reason, we perform first the combination with data taken from the NVSS. These data have lower sensitivity than the VLA ones, but cover a wider area filling the whole region observed with SRT. This fact has two advantages: firstly,  
extending the region covered by both interferometric and single dish instruments, 
it allows to have a higher number of sources present in both single dish and interferometric data sets, increasing in this way the statistic of fluxes comparison, and therefore providing a more robust estimate of the scaling factor between the two data sets.
Secondly, it allows to perform the combination 
also at large distance from the cluster, 
where, as we have seen, there are some possibly interesting structures.
Therefore we show first the procedure and the results for the combination with the NVSS data, and later the results for the combination with the pointed VLA data.

\subsection{Combination with NVSS data}

We describe here the procedure for combining the image obtained with SRT with the 
NVSS one, in order to obtain information on both large and small angular scales. For this purpose, we followed the procedure detailed in Loi et al. (2017), Vacca et al. (2018), and Murgia et al. (2024), which is based on a combination of the two images in the two-dimensional Fourier space, and on the calculation of the factor that allows to have a comparable power in the Fourier spectra of the two images in the range of the angular scales where both the images are sensitive. 

In order to perform the combination, it is necessary that the single-dish and the interferometric images are selected in the same frequency band and region in the sky, because we need to directly compare their powers for determining the correct scaling factor. Therefore
we first obtained an SRT image of the cluster in the same frequency bands of the NVSS (IF1: 1343.9 -- 1385.9 MHz; IF2: 1414.1 -- 1456.1 MHz), and 
determined the properties of the SRT beam
by performing a fit to the image of the flux calibrator 3C147 in these bands, obtaining as a result
a circular beam with FWHM axis of 835 arcsec (i.e. $\sim13.9$\,arcmin), while the NVSS has a
circular beam with FWHM axis of 45 arcsec. 
We gridded the NVSS and SRT images with the same spacing, and blanked the two images in order to include exactly the same area of the sky, in order to ensure that we are considering the same range of angular scales.

Later, we 
calculated the NVSS and SRT Fourier spectra and deconvolved the results by dividing by the Fourier transform of the respective beam.
After this step, we calculated the multiplicative factor for which the two spectra overlap each other at angular scales between a minimum value, determined from the SRT resolution, and maximum value, determined from the maximum scale at which the interferometric measures of NVSS are sensitive. 
By inspecting the Fourier spectra and searching for the overlapping region, we determined the optimal values for these angular scales 
to be 14.8 and 20.6 arcmin respectively. The multiplicative factor for which the SRT needs to be scaled in order to make the powers of the two spectra comparable resulted to be $0.942\pm0.082$, where the error bars is estimated from the dispersion of the SRT and NVSS power distributions in the Fourier plane.

Once the scaling factor has been determined, we merged the two spectra assuming different weights for the 
different regions of the Fourier plane: the weights are 1 for SRT and 0 for NVSS at the longest scales ($>20.6$ arcmin), 0 and 1 at the shortest scales ($<14.8$ arcmin), and vary linearly from 1 to 0 for SRT (and from 0 to 1 for NVSS) in the overlapping region. The merged spectrum is then convolved with the NVSS beam, by multiplying it by the corresponding Fourier transform, and finally the combined image is obtained by performing the inverse Fourier transform of the resulting spectrum.
 
In the top-left panel of Fig.\ref{fig.confronto_radio} we show the spatial scale spectra of the NVSS and SRT images at the various stages of the procedure,
where the overlapping region between the two spectra is highlighted with the vertical dashed lines.

In order to appreciate 
the contribution coming from the SRT data to the combined image,
in the bottom-left panel of Fig.\,\ref{fig.confronto_radio} we show the difference between the combined SRT+NVSS image and the NVSS one, overlapped with the NVSS contours. It is particularly evident, approximately around the position (02:18:00 +70:20:00), a structure located at the south-east of the cluster, which is instead at the center of the image.

\subsection{Combination with VLA data}

After the combination with NVSS, we combined the SRT image with the image obtained from VLA pointed observations, working in the bands 1372.6--1410.1\,MHz and 1452.4--1489.9\,MHz.

In order to perform the combination, we followed the same procedure as in the case of the combination with the NVSS data, determining first the SRT beam in the VLA band by performing a fit to the calibrator 3C147; the SRT beam resulted to be circular, with a FWHM axis of 826 arcsec (i.e. $\sim13.8$\,arcmin).
In the combination, we used as scaling factor for the SRT data the same value found when we performed the combination with NVSS data, i.e. 0.942; we made this assumption because the NVSS data cover a wider area than the pointed VLA ones, so, as previously discussed, this value has been obtained using a better statistics for calculating the angular power spectrum than the one obtainable using VLA data. We also checked that, if we had calculated the scaling factor from the comparison between VLA and SRT power spectra, we had found $0.81\pm0.15$, i.e. a value compatible within 1$\sigma$ with the one obtained from NVSS, but with a higher relative error (19\% instead of 9\%).

\begin{figure*}
\centering
\begin{tabular}{c}
\includegraphics[width=0.5\textwidth, trim={0 5cm 0 3cm}, clip]{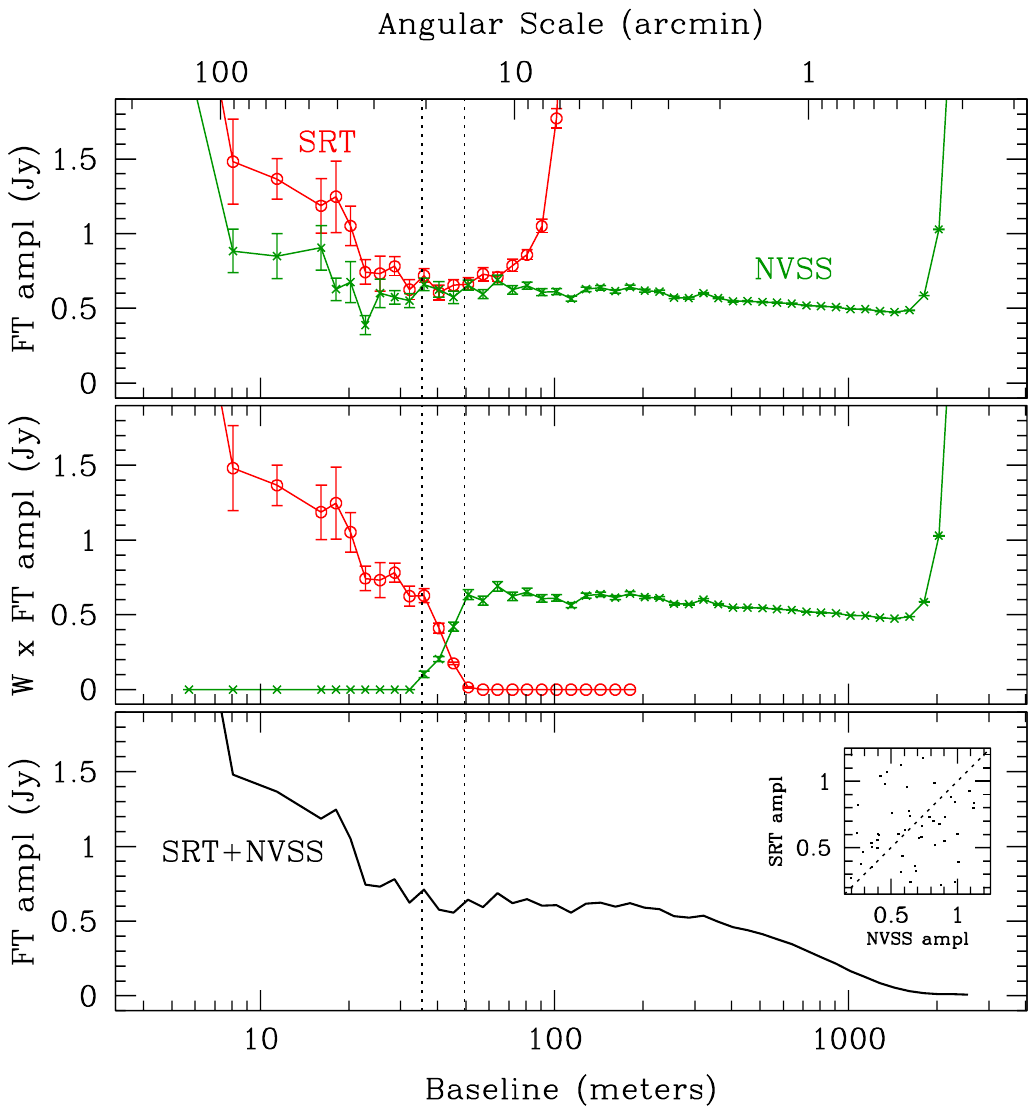}
\includegraphics[width=0.5\textwidth, trim={0 5cm 0 3cm}, clip]{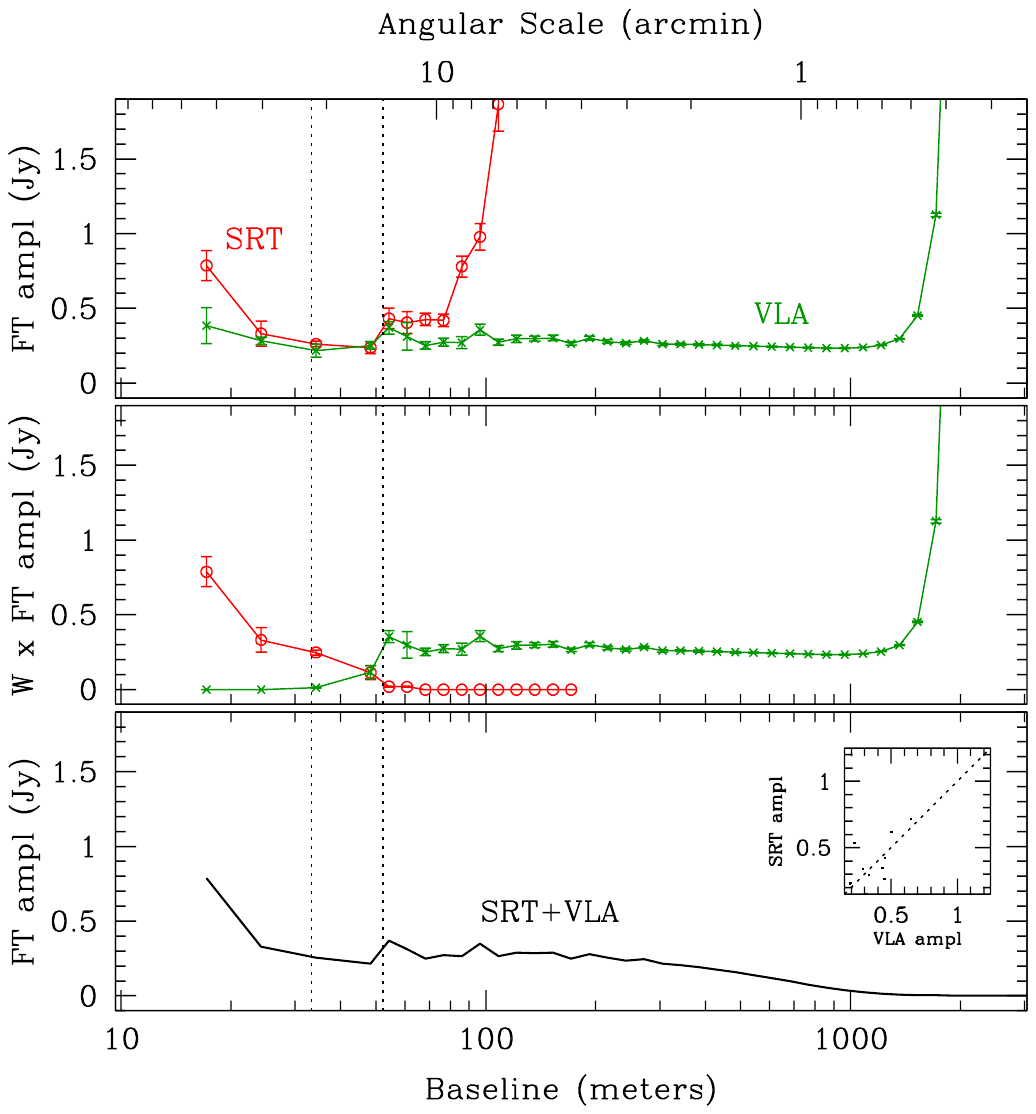}\\
\includegraphics[width=0.5\textwidth, trim={1cm 9cm 0.5cm 1cm}, clip]{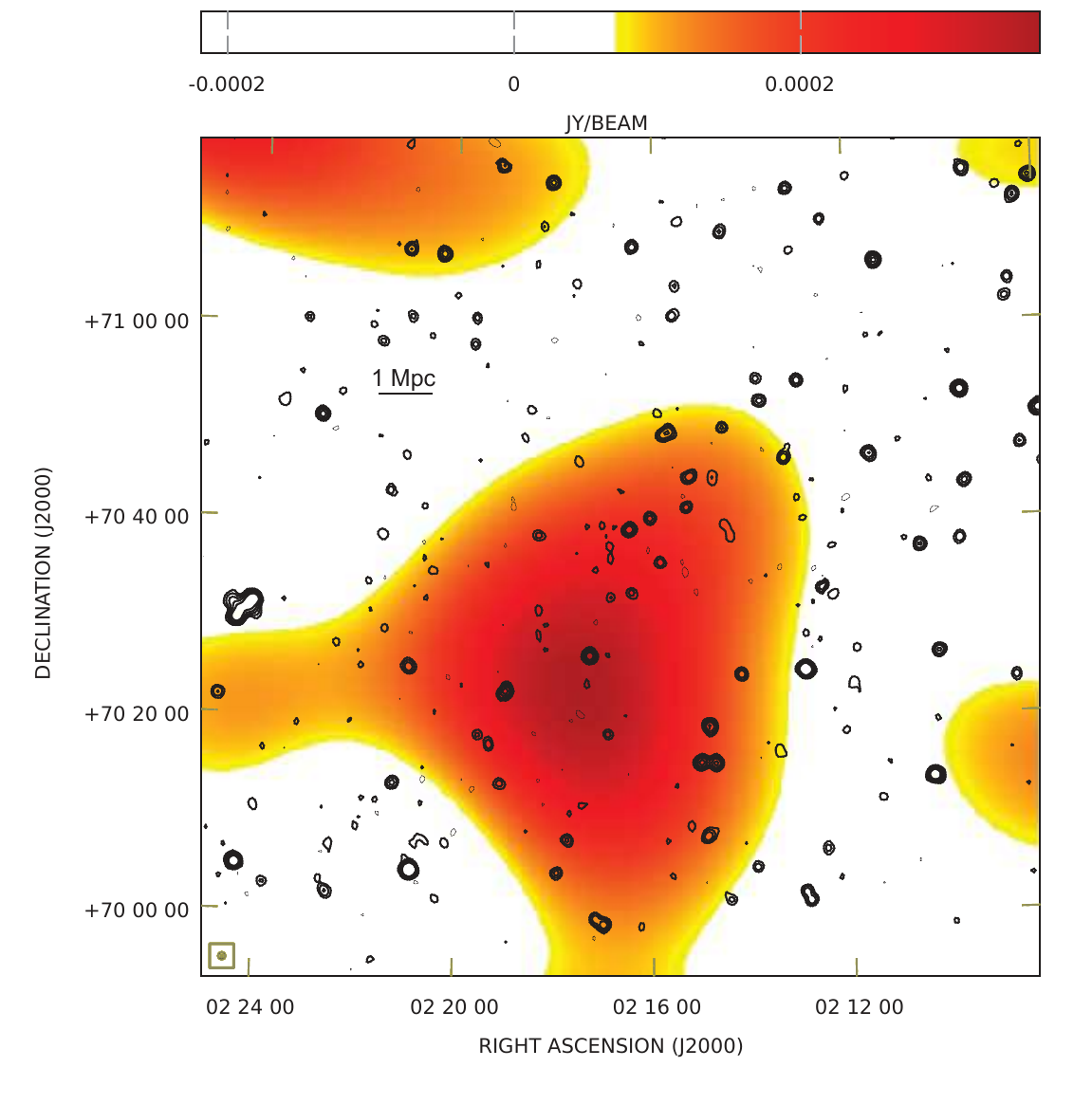}
\includegraphics[width=0.5\textwidth, trim={1cm 9cm 0.5cm 1cm}, clip]{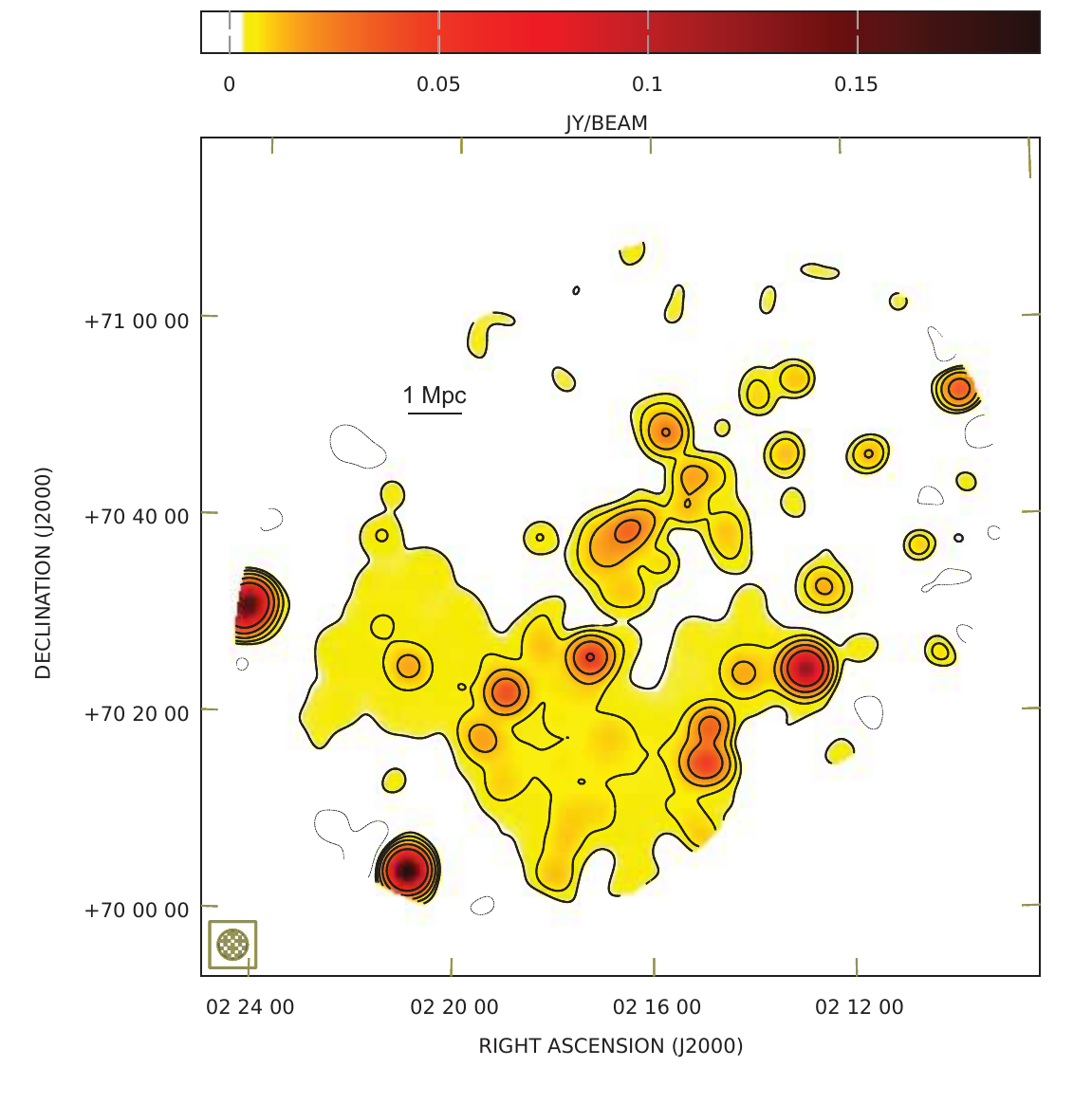}
\end{tabular}
\caption{
\textit{Top panels:} Fourier spatial scale spectra of SRT and NVSS data (left) 
and of SRT and VLA data (right) used for performing the combination. 
In these plots of the Fourier spectra, in the top panel the separated power spectra of the SRT data (red) and the NVSS/VLA data (green), deconvolved by the respective beams and before the alignment, are shown.
In the intermediate panel, the weighted power spectra of the SRT data (red) and the NVSS/VLA data (green) before the flux density alignment are shown. 
In the bottom panel the combined power spectrum after the alignment tapered with the beam of the interferometer is shown. In the inset, the comparison between the amplitude of the SRT and of the NVSS/VLA data in the region of overlapping in the Fourier space before the alignment is shown.
The dashed vertical black lines identify the overlapping region used to adjust the flux density scales. 
\textit{Bottom-left panel:} Image obtained by subtracting the NVSS image from the combined SRT+NVSS image; NVSS contours at levels of $-3\sigma$ (dotted),  $3\sigma$, and scaling with a factor of two (solid), with $\sigma=0.45$ mJy/beam, are overlapped. The NVSS beam (FWHM 45 arcsec) is shown in the bottom-left corner of the image, and the horizontal bar indicates a length of 1\,Mpc at the cluster distance.
\textit{Bottom-right panel:} Combined SRT+VLA image convolved with a 3 arcmin beam (shown in the bottom-left corner of the image); contours are at levels of $-3\sigma$ (dotted), $3\sigma$, and scaling with a factor of two (solid), with $\sigma=1$ mJy/beam. The horizontal bar indicates a length of 1\,Mpc at the cluster distance.}
\label{fig.confronto_radio}
\end{figure*}

We estimated the values of the minimum and maximum angular scales where the single dish and the interferometric data spectra overlap each other, 
assuming first the values we can expect based on their baseline ranges, and later refining them through close inspection of the spectra, obtaining
the values of 14 and 22 arcmin respectively. 
In the top-right panel of Fig.\,\ref{fig.confronto_radio} we show the Fourier spectra of the SRT and the VLA images, highlighting the region of the angular scales where there is the overlapping of the two spectra. 
In the bottom-right panel of Fig.\,\ref{fig.confronto_radio} we show the combined SRT+VLA image 
convolved with a 3 arcmin beam, in order to better appreciate the extended features.

\begin{figure}
\centering
\begin{tabular}{c}
\includegraphics[width=0.5\textwidth, trim={1.5cm 9.5cm 0 0.7cm}, clip]{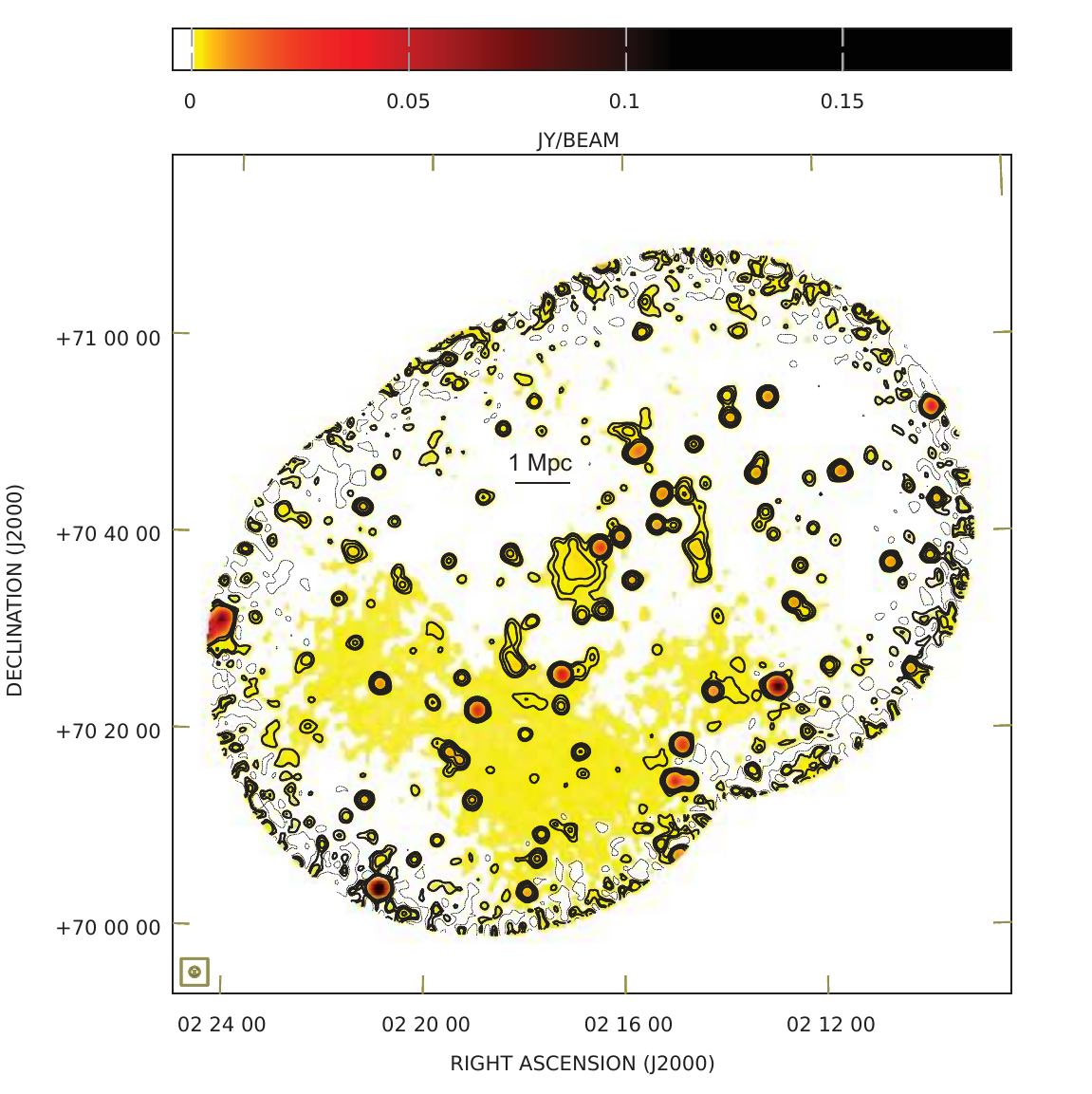}
\end{tabular}
\caption{Combined SRT+VLA image overlapped with VLA contours at levels of $-3\sigma$ (dotted),  $3\sigma$, and scaling with a factor of two (solid), with $\sigma=0.15$ mJy/beam. Colors are saturated at the corresponding $3\sigma$. The FWHM beam of the image, indicated in the bottom-left corner, is 61 arcsec.}
\label{fig.combo_vla}
\end{figure}

In order to point out the importance of combining single dish and interferometric images, we compare in Fig.\,\ref{fig.combo_vla} the combined SRT+VLA image (colors) with the contours from the VLA image; it is evident that the combined image shows the presence of large scale emission not visible in the VLA-only image, in particular in the region in the south-east of the cluster. In the next section we will discuss about this emission.

\section{Discussion}

In this Section we inspect the images we have obtained, i.e. the SRT image (contours in left panel of Fig.\,\ref{fig.radio_Xray}) and the combined SRT+VLA image (bottom-right panel of Fig.\,\ref{fig.confronto_radio} and Fig.\,\ref{fig.combo_vla}). These images show the presence, in addition to the already known diffuse emissions belonging to the cluster (halo and relics), of several large scale extended structures located outside the cluster. In particular, it is evident a large extended structure at the south-east of the cluster, to which the cluster is possibly connected. 

We divide our discussion in two parts: first, we study the properties of the diffuse emissions detected in the cluster, in particular of the radio halo, and compare our results to the ones obtained by Hoang et al. (2021), in order to appreciate how much additional information can be obtained by combining single dish and interferometric observations. Later, we describe the properties of the south-eastern extended emission, attempting to derive some information about its nature.

\subsection{Properties of the radio halo}

We present here the results of the study of the region of the radio halo. We 
estimate the halo flux density by 
performing a two-dimensional exponential fit to the azimuthally averaged brightness profile calculated in concentric annuli centered on the cluster center. 
Moreover, we derive a map of the spectral index spatial distribution and the halo integrated spectral index between the frequencies of 140 MHz and 1.4 GHz, by comparing the combined image obtained in this paper with a publicly available LOFAR image.

\subsubsection{Flux density}

The presence of a diffuse radio halo centered at the cluster center, and having a size of the order of $\sim6$ arcmin ($\sim1$\,Mpc at the cluster distance), is clearly visible in the VLA image (right panel of Fig.\,\ref{fig.radio_Xray}) and in the SRT+VLA combined image (Figs.\,\ref{fig.confronto_radio} and \ref{fig.combo_vla}). 

In order to derive quantitative information on the spatial shape of the radio halo and an estimate of the total flux density,  
we performed an exponential fit to its azimuthally averaged radial profile. In this way we can obtain a result that is not biased by the sensitivity of the image, unlike the result that could be obtained by integrating the surface brightness above a $3\sigma$ threshold.

We first calculated the average surface brightness inside seven annuli centered on the peak of the RASS image of the cluster, and with a spacing of half of the VLA beam (30.5 arcsec), verifying that the average surface brightness calculated in an eighth annulus is lower than the $3\sigma$ level of the image, and therefore only an upper limit can be derived in this last annulus. When doing this operation, we masked the discrete sources (identified from a higher resolution map, see fig.\,A.1 in Hoang et al. 2021), and assumed that the masked pixels have a surface brightness equal to the average one calculated inside the corresponding annulus (see Fig.\,\ref{fig.exp_masks}). 

We assumed that the error bar of the surface brightness calculated in each annulus can not be smaller than $10\%$ of the average surface brightness in that annulus; 
in this way, we take into account the presence of possible systematic errors,  
due for example to the way the masked regions are defined or to deviation from the azimuthal symmetry, and can avoid that annuli with more solid statistic, i.e. the external ones where the number of pixels is higher, would have a too strong weight on the fit result compared to other annuli; other possible sources of systematic errors are the uncertainties in the calibration flux scale (Perley \& Butler 2017) or, for the combined image, on the single dish scaling factor.

\begin{figure}
\centering
\begin{tabular}{c}
\includegraphics[width=\columnwidth, trim={0 5cm 0 4cm}, clip]{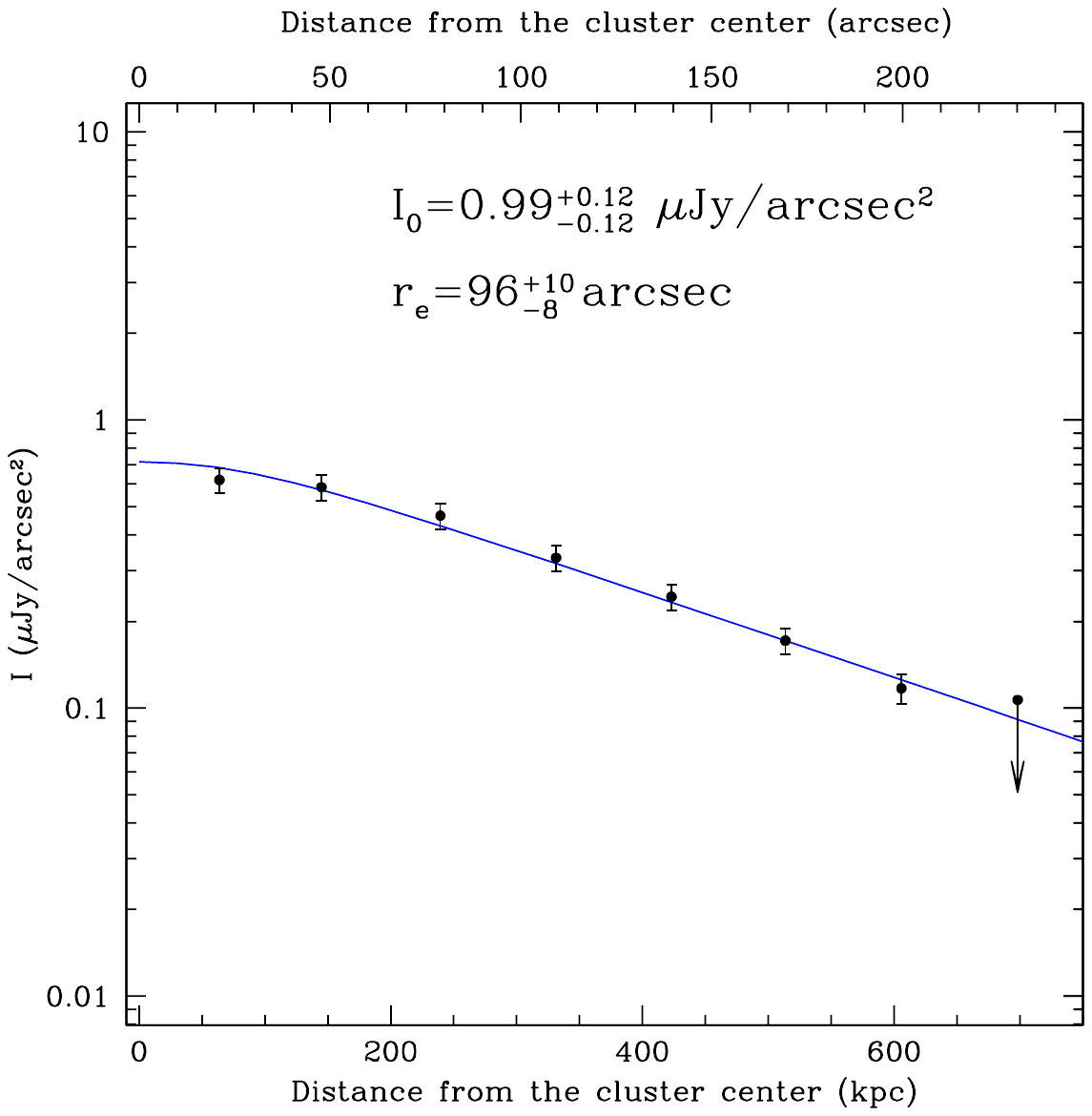}
\end{tabular}
\caption{Surface brightness azimuthally averaged radial profile of the combined SRT+VLA image of the radio halo fitted with an exponential function of the type $I=I_0 \exp (-r/r_e)$.}
\label{fig.halo_exp_profile}
\end{figure}

Once we have obtained the average surface brightness in each annulus, we fitted the radial profile using a two-dimensional exponential function of the type $I=I_0 \exp (-r/r_e)$ (Murgia et al. 2009). 
For the VLA image we obtained $I_0=(0.96^{+0.13}_{-0.13})\,\mu$Jy arcsec$^{-2}$ and $r_e=91^{+11}_{-10}$ arcsec, while for the combined image we obtained $I_0=(0.99^{+0.12}_{-0.12})\,\mu$Jy arcsec$^{-2}$ and $r_e=96^{+10}_{-8}$ arcsec (see Fig.\,\ref{fig.halo_exp_profile}).
The result of the fitting procedure is statistically good: we obtain, for the combined image, a value of the reduced $\chi^2$ of 0.459 with 5 d.o.f.

The total flux densities derived by integrating these profiles up to $3r_e$ are $F=(40.1^{+6.0}_{-4.8})$\,mJy for the VLA image, and $F=(45.6^{+5.3}_{-4.6})$\,mJy for the combined image. 
We note that by employing this method to estimate the halo flux density there is some arbitrariness in establishing the limit of integration at the distance of $3r_e$ (see Murgia et al. 2009 for details); this fact, as well as other possible sources of systematic errors, including, as previously mentioned, possible deviations from the spherical symmetry, are taken into account when establishing the error bars in each annulus to be not lower than 10\% of the average surface brightness in that annulus.

By considering the best fit values without considering the error bars,
we can note that the flux density in the combined image is higher by a factor of $\sim14\%$ compared to the VLA-only image; since the central value $I_0$ of the combined image is higher by a factor of only $\sim3\%$, we can conclude that the main source of difference is 
the larger extension of the halo in the combined image, described by the different best fit values of $r_e$. 

However, considering the error bars, the results found in the two images are compatible; this means that the combination procedure in this cluster
can in principle allow to recover a fraction of the flux density that is different from zero, but this difference is not statistically significant at $1\sigma$ level. 
Anyway, we note that, by applying the same method to derive the total flux density in the VLA and in the combined image, the combination procedure can provide a higher flux only if there is actually emission on scales larger than those detectable by the interferometer; for this reason, we believe that the correct value is that of the combined image.

We can compare our results with the one found with an exponential fit by Hoang et al. (2021), $F=(58.3\pm3.4)$\,mJy:
we note that their result was calculated up to $4r_e$ instead of $3r_e$ as in our case; rescaling their value to a maximum radius of $3r_e$ we obtain a flux density of $F=(51.3\pm3.0)$\,mJy; this value is
higher than the values we found in both the VLA-only and the combined image, and compatible within $1\sigma$ only with the combined one.
This discrepancy could arise from differences in the fitting procedure or in the subtraction of discrete sources.

By comparing the value of $r_e$ obtained by Hoang et al. (2021), $r_e=281\pm7$\,kpc, with the ones we found, $r_e=276^{+33}_{-30}$\,kpc for the VLA-only image and $r_e=291^{+30}_{-24}$\,kpc for the combined one, we note that 
our result for the VLA image is very similar, while the result for the combined image is slightly higher, but compatible inside the error bars.

\begin{figure}
\centering
\begin{tabular}{c}
\includegraphics[width=\columnwidth, trim={1.5cm 6.5cm 2cm 7cm}, clip]{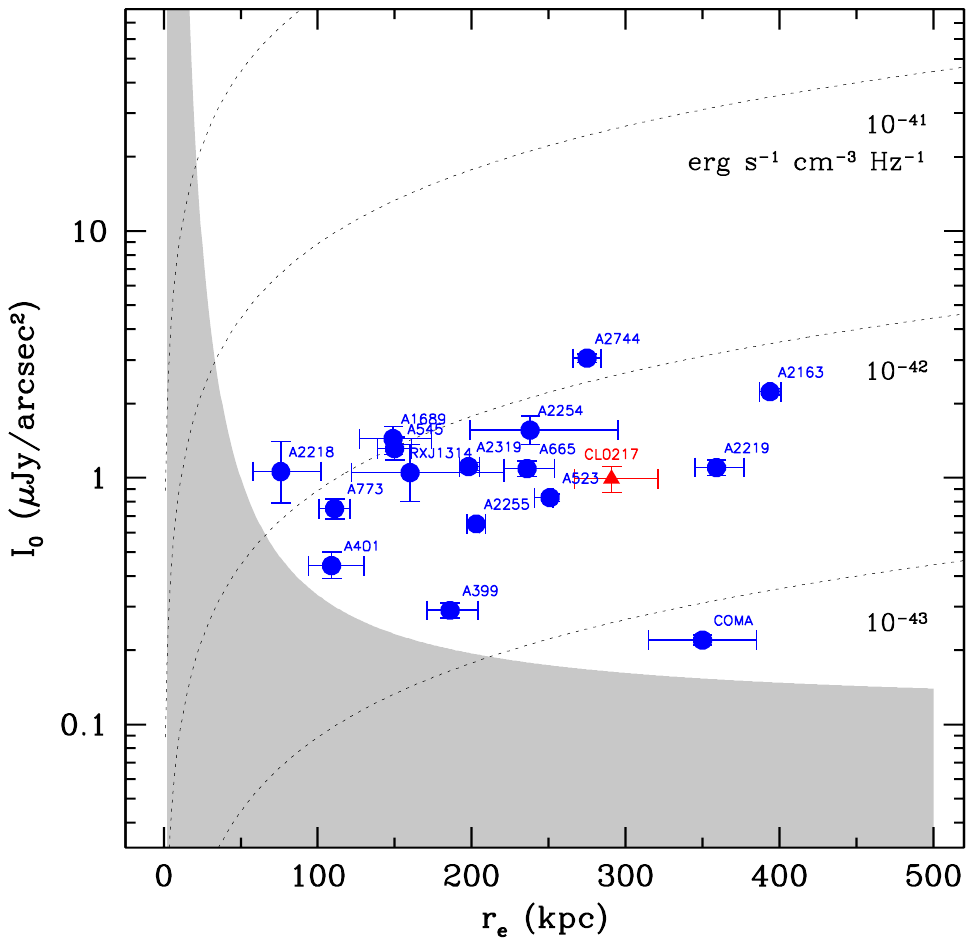}
\end{tabular}
\caption{$I_0-r_e$ plane at 1.4 GHz for a sample of known radio halos, including the values for the cluster CL0217 as obtained in this paper. Radio emissivity is constant along the dotted lines. The gray area represents the undetectability region for a radio halo at redshift $z>1$ observed at a resolution of 60 arcsec due to the confusion noise (see Murgia et al. 2024 for details).}
\label{fig.halo_exp_plane}
\end{figure}

In Fig.\,\ref{fig.halo_exp_plane} we compare the values of $I_0$ and $r_e$ obtained in this paper for the combined image 
with the values found in other clusters (Murgia et al. 2009, 2010; Vacca et al. 2011; Murgia et al. 2024); we can see that the radio halo in CL0217 has one of the largest known linear sizes, and a central surface brightness which instead is typical of other halos in the intermediate part of the clusters distribution, with an emissivity comparable to the one of other clusters like A2219, A523, and A2255.

\subsubsection{Spectral index}

We estimated the spectral index of the radio halo using our combined SRT+VLA image and a publicly available LOFAR image (Hoang et al. 2021\footnote{image downloaded from https://vizier.cds.unistra.fr/viz-bin/VizieR?-source=J/A+A/656/A154 selecting the one with uvrange 0.12-61 k$\lambda$, robust 0.25, taper 25 arcsec}). We convolved the LOFAR image to the same beam of the VLA one (61 arcsec), and gridded the two images in order to have the same cell size (4 arcsec/pixel as the original LOFAR image) and a 1:1 pixel by pixel match. Hence, we derived the integrated spectral index between the two frequencies of the LOFAR and SRT+VLA images (140 MHz and 1.4 GHz respectively), calculating the flux densities at the two frequencies by integrating over the pixels having surface brightness larger than $3\sigma$ at both the frequencies inside the halo region; 
in this operation we masked the discrete sources as in Fig.\,\ref{fig.exp_masks}, assuming that the surface brightness in the masked pixels is equal to the average calculated in the halo. 
We calculated the error bar on the flux density as the image noise multiplied by the number of beams contained in the halo, adding in quadrature 
a systematic error due to uncertainties on the flux scale of $20\%$ of the flux for the LOFAR image, and of $5\%$ for the SRT+VLA one. We obtained a value of the integrated spectral index of $\alpha=1.060\pm0.089$. This value is in accordance with the value found by Hoang et al. (2021), i.e. $1.07\pm0.05$, obtained by integrating over the regions with surface brightness larger than $2\sigma$.

\begin{figure*}
\centering
\begin{tabular}{c}
\includegraphics[width=\textwidth, trim={1.2cm 19cm 0.6cm 0.5cm}, clip]{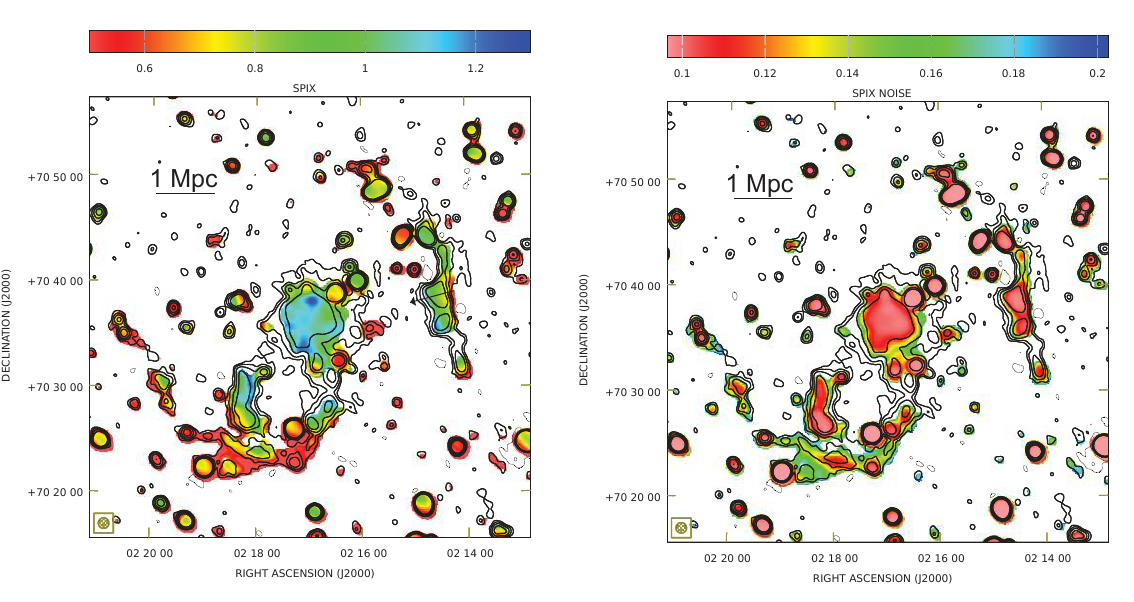}
\end{tabular}
\caption{Spectral index map (left panel) and error on the spectral index (right panel) between 140 MHz and 1.4 GHz calculated using a LOFAR image and the combined SRT+VLA image obtained in this paper. Contours are taken from the LOFAR image, and are at levels of $-3\sigma$ (dotted),  
$3\sigma$ and scale with a factor of two (solid), with $\sigma=0.3$ mJy/beam. The beam of 61 arcsec is shown in the bottom-left corners and the horizontal bars indicate a length of 1 Mpc at the cluster distance. The arrow in the left panel indicates the direction across the north-western relic where the spectral index is progressively steepening inwards.}
\label{fig.spectral_index}
\end{figure*}

\begin{figure}
\centering
\begin{tabular}{c}
\includegraphics[width=\columnwidth, trim={0 5cm 0 4cm}, clip]{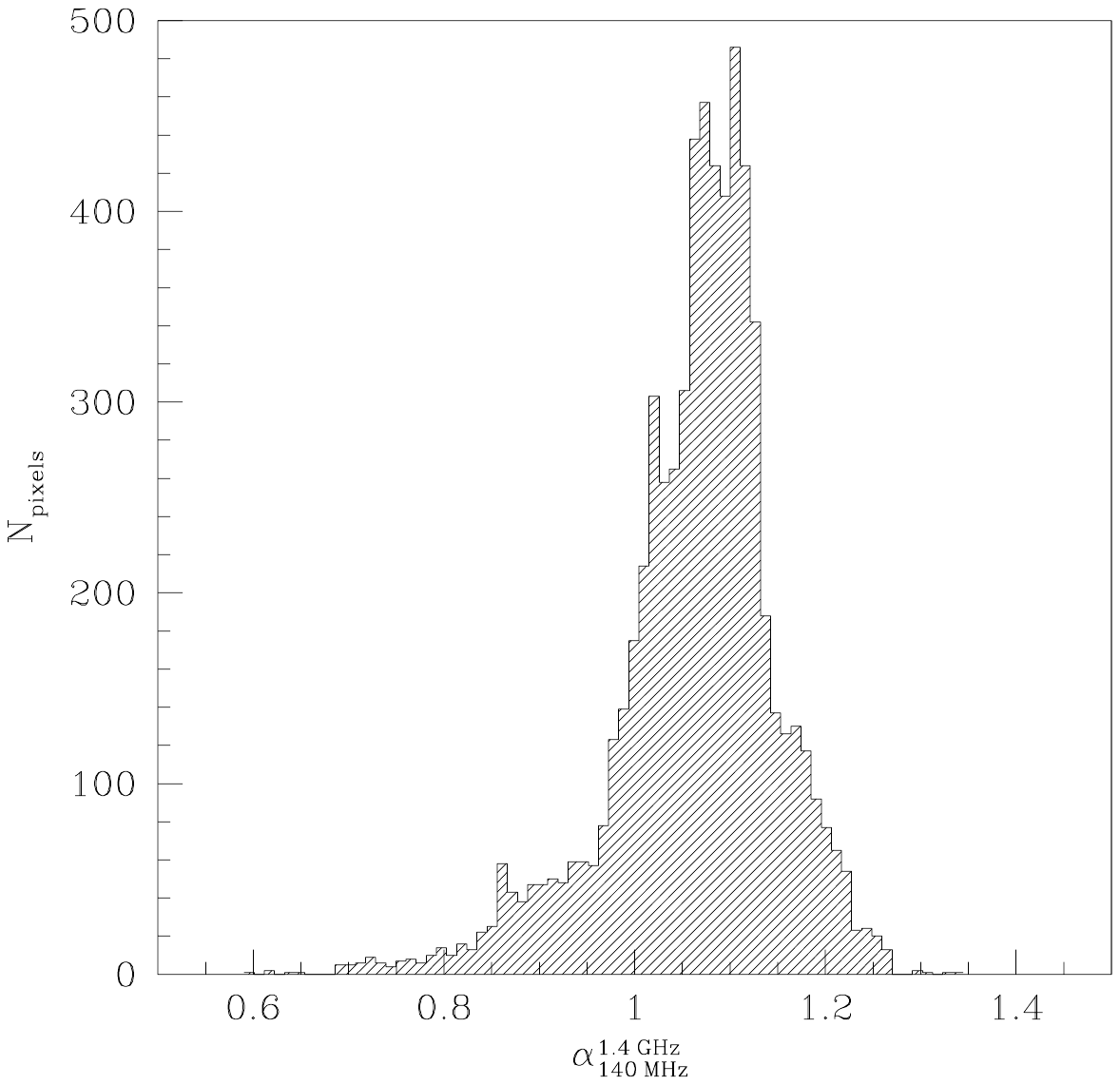}
\end{tabular}
\caption{Histogram of the distribution of the spectral index inside the halo. This distribution has a mean value of 1.064 and a rms of 0.086.}
\label{fig.spix_histo}
\end{figure}

We also obtained a spectral index map between the same frequencies, which is reported in Fig.\,\ref{fig.spectral_index}. Comparing this map with the analogue one published in fig.\,6 in Hoang et al. (2021), we can see that the results are in accordance for the halo and the relics. In the halo the spectral index is quite uniform, with fluctuations in the range $\sim1-1.2$, and with a flatter part in the northern region, close to the northern X-rays discontinuity detected by Zhang et al. (2020); excluding this part of the halo, the distribution of the spectral index values results to have a mean value of 1.064 with a rms of 0.086 (see Fig.\,\ref{fig.spix_histo}). 
In the north-western relic we observe a typical spectral index structure, with the flattest part located along the external edge, and with a progressive steepening in the direction of the cluster center (highlighted with an arrow in the left panel of Fig.\,\ref{fig.spectral_index}).

An evident difference between our spectral index map and the one of Hoang et al. (2021) can instead be seen in the outer region of the south-eastern relic: our map shows the presence of a region with a relatively flat value of the spectral index ($\alpha\sim0.6$) extending beyond the edge of the relic; this part of the image was not detected in the spectral index map presented by Hoang et al. (2021). It is possible that the combination with the SRT data has allowed us to recover some large scale structure at 1.4 GHz that was not detected in the VLA-only image (probably because it was under the $3\sigma$ level, unlike in the LOFAR image), therefore allowing to detect a wide region where the spectral index is quite flat and uniform. The presence of this region might challenge the explanation of the spectral index structure in this region as due to shock acceleration, because in this case the flattest value of the spectral index would be expected just in a narrow region, as observed in many relics, including the north-western relic in this cluster.

Therefore it is not clear if this flat spectrum region is associated with the relic or with some other extended emission.  
It might be associated with the discrete source located at the south-eastern corner of this region. Alternatively, it can be part of the large extended structure located at the south-east of the cluster, as we discuss in the next subsection, or can have a Galactic origin.

Finally, we note that our estimates of the spectral index may be affected by the fact that we used two images obtained with LOFAR and VLA that have been obtained with different values of the interferometric settings, as uv range, robust, and tapering; moreover, the sampling of the central part of the uv-plane is different in the two cases. These facts may have affected in some way our estimates of surface brightness and flux, and therefore of the spectral index. However, we note that we have convolved the two images with the same beam of 61 arcsec, and therefore any information about small angular scales (which in principle is different in the two images because of the different value of the maximum uv coverage) has been eliminated, reducing this kind of bias in our estimates. The fact that our spectral index map in the cluster region is in accordance with the one presented in Hoang et al. (2021), who instead have produced the map by selecting the same sampling of the uv plane, suggests us that the bias present in our results should be negligible (for a similar case see, e.g., Boschin et al. 2023).  

\subsection{Properties of the south-eastern extended emission}

As already noted, the combined SRT+VLA image shows the presence in its south-east part of a wide region with lower surface brightness compared to the central regions of the cluster, but large enough to be above the $3\sigma$ level in a wide area. This region, as visible in Fig.\,\ref{fig.combo_vla}, appears to have a ``V'' shape, or an inclined ``L'' shape with the longest arm extending from the east of the cluster up to its south, and the shortest (and less bright) arm extending up to the west of the cluster.

This structure does not seem to coincide with any extended X-ray structure, and therefore is probably not associated with other clusters (or sub-clusters) interacting with CL0217. Its angular extension is very large compared to the cluster, with the longer side being of the order of 50\,arcmin, which, if located at the cluster distance, would correspond to a size of the order of 10\,Mpc, making it unlikely to be related to the cluster. On the other hand, the south-eastern cluster relic appears to be embedded inside part of this structure, suggesting that these structures may be instead related. 

As visible in Fig.\,\ref{fig.combo_vla}, this region, especially in its southern part, extends very close to the edge of the region covered by VLA data, where the detected surface brightness can be artificially high because of the primary beam correction. Since the characterization of the VLA beam in the L-band has shown that the power response of the instrument is of the order of 0.5 when the offset from the center of the pointing is $\sim20$ arcmin (EVLA Memo 195, Perley 2016), we have blanked all the pixels in the image that are not at a distance of less of 20 arcmin in at least one of the three pointings that have been used in this paper. In this way some parts of the image close to the edges have been excluded from this analysis.

We measured the total flux density of this region, dividing it in two rectangular regions as shown in Fig.\,\ref{fig.se_masks}, and masking all the discrete sources visible in the combined image, as well as a circular region centered outside the rectangular one, but with its south-eastern part that coincides with the cluster relic, which seems to be embedded in the extended emission. Considering the pixels having surface brightness higher than $3\sigma$, we measured the average surface brightness of the two rectangular regions, obtaining the values of $\langl I \rangl=(0.64\pm0.18)$\,mJy/beam for the larger one, and $\langl I \rangl=(0.66\pm0.26)$\,mJy/beam for the smaller one. 

In order to calculate the total flux density, we assumed that the pixels corresponding to the masked sources embedded in the extended emission, excluding the relic and the five sources located at the extreme northern part of the larger region, as well as the two sources located in the northern part of the smaller region, have a surface brightness equal to the average one calculated in the corresponding rectangle. We have calculated the error bar on the flux density as the image noise multiplied by the number of beams contained in the area, and by adding in quadrature a component of $5\%$ of the flux to take into account the uncertainties in the flux density scale (Perley \& Butler 2017). With these assumptions, we obtained flux density values of $F=(190.2\pm9.9)$\,mJy for the larger region, and $F=(46.4\pm2.6)$\,mJy for the smaller one. By summing these two values, we obtain a total flux density for the whole region of $F=(237\pm10)$\,mJy.

It is not immediate to understand if this extended emission is located at the distance of the cluster (as in the case of a large filament) or at a closer distance, like in the case of a Galactic foreground emission, which is a probable explanation given the low latitude of the cluster, $b\sim8^\circ$. 
If it is located at the cluster distance, we can estimate its emissivity in order to understand if it is typical of large scale filaments, and therefore if this kind of origin can be excluded.

In order to derive an estimate of the emissivity, it is necessary to make an assumption about the volume of the region. For this purpose, we estimated that the two rectangular regions, if located at the cluster distance, would have areas of approximately 23 and 9.8 Mpc$^2$ respectively. Assuming that these regions have an extension along the line of sight equal to the smallest of their sides, i.e. $\sim3.1$\,Mpc, we obtain a total volume of both these regions of $V=102$ Mpc$^3$. Given the distance of the cluster of $D_L=872$ Mpc, we estimated the emissivity as $J=4\pi D_L^2 F/V$, obtaining a value of $7.2\times10^{-44}$ erg cm$^{-3}$ s$^{-1}$ Hz$^{-1}$ (we note we did not apply any k-correction because the spectral index of the source is not known). This value is of the order of magnitude of the weakest among observed candidate filaments connecting galaxy clusters on large scale (Vacca et al. 2018; Govoni et al. 2019), and therefore it can not be excluded that also this emission has a similar nature on the basis of its estimated emissivity. On the other hand, its very large size and the fact that it does not seem to point towards the only cluster observed in the region, seem to disfavor this possibility.

We also inspected the LOFAR image by searching for some extended emission in this region, but we did not find any hint of it.
We note however that the feature is largely outside of the LOFAR field. For the area present in the LOFAR image, using the 1.4 GHz brightness from the SRT+VLA image and the $3\sigma$ limit in the images from Hoang et al. (2021), we obtain an approximate upper limit to the spectral index $\alpha < 1.4$, which is rather inconclusive.

\section{Conclusions}

In this paper we have presented the results of observations of the galaxy cluster CL0217 and its surrounding region performed with SRT in the 1.3--1.8 GHz band. As expected, the SRT angular resolution in this frequency range is not sufficient to resolve the internal structure of the cluster; on the other hand, since SRT is a single dish telescope, the resulting image is sensitive up to the largest scales probed inside its field of view, which has an area of $3^\circ \times 3^\circ$ in this study.

In order to obtain information also on smaller scales, we have analyzed publicly available archival VLA data of this cluster, obtaining a detailed image of the internal structure of the cluster, showing the structures that were already known from literature, as the central radio halo, a north-western relic, and a south-eastern possible relic (Brown et al. 2011; Hoang et al. 2021).

We have combined the single dish and the interferometric data following a procedure based on the alignment of the single dish flux scale with the interferometric one and the merging of the Fourier spectra (Loi et al. 2017; Vacca et al. 2018), obtaining a combined image having the angular resolution of VLA but sensitive up to largest scales as the SRT one.

By measuring the flux density of the radio halo, we have found that the value obtained for the combined SRT+VLA image is higher than the one obtained for the VLA-only image by a factor of $\sim14\%$, even if this difference is smaller than the error bars, therefore not significant at $1\sigma$ level. This result has been obtained  
by performing a two-dimensional fit to the azimuthally averaged surface brightness profile calculated in concentric annuli centered on the cluster center. Interestingly, 
while the derived average values of the central surface brightness are quite similar in the SRT+VLA and in the VLA-only image, the main difference in the flux density is due to a larger extension of the radio halo in the combined image. This suggests that the combination procedure did not introduce any spurious flux in the image, but just allowed us to recover a higher fraction of the flux on large scale.

We also derived a spectral index map comparing the SRT+VLA image with a publicly available LOFAR map at 140 MHz from Hoang et al. (2021). By comparing our result with the spectral index map published by Hoang et al. (2021), which was obtained comparing the LOFAR with the VLA-only map, we obtain a very similar result, with a radio halo presenting a quite uniform distribution of the spectral index, and a north-western relic presenting the flattest values of the spectral index in a thin external region, and a progressive steepening inwards. There is instead a difference in the external part of the south-eastern candidate relic, where there is the presence of a relatively large region having a spectral index equal to the flattest one found in this relic ($\alpha\sim0.6$). This region can challenge the real nature of this relic, because typically the flattest spectral index in a relic is expected to be observed only in a thin region, where a shock front is accelerating the electrons. It is therefore unclear whether this region is a relic seen from an unusual angle of view, or is the combination of more than one relic, or something else, such as the tail of a radio galaxy, or is part of another structure external to the cluster.

The south-eastern candidate relic is in fact possibly embedded inside an interesting extended structure visible in the combined SRT+VLA image. This structure is located outside the cluster in the south-eastern area, has an inclined L-shape, and extends along its larger side for about 50 arcmin. If located at the distance of the cluster, this extension would correspond to a linear length of about 10 Mpc. We have derived the flux density of this structure, and, under the hypothesis it is located at the cluster distance and its extension along the line of sight is equal to the smaller projected side, about 3 Mpc, we calculated its emissivity, which resulted to be of the order of magnitude of other candidate large scale filaments connecting clusters (Vacca et al. 2018). On the other hand, this structure does not point towards the cluster, making unlikely it is a cosmic web filament. 
An alternative possibility is that the south-eastern extended structure is a foreground emission originated in our Galaxy; this possibility is favored by the fact that this cluster is located at low Galactic latitude. New observations of this structure, aimed at better determining some properties like its spectral index, would be useful to understand if its origin is Galactic or extragalactic.

\section*{Acknowledgments}

The Enhancement of the Sardinia Radio Telescope (SRT) for the study of the Universe at high radio frequencies is financially supported by
the National Operative Program (Programma Operativo Nazionale - PON) of the Italian Ministry of University and Research ``Research and
Innovation 2014-2020'', Notice D.D. 424 of 28/02/2018 for the granting of funding aimed at strengthening research infrastructures, in
implementation of the Action II.1 -- Project Proposals PIR01$\_$00010 and CIR01$\_$00010. 
VV acknowledges support from the Prize for Young Researchers ``Gianni Tofani'' second edition, promoted by INAF-Osservatorio Astrofisico di Arcetri (DD n. 84/2023).
The authors thank C. Riseley and the Referee for useful comments and suggestions.

\section*{Data availability}

The data underlying this article will be shared on reasonable request to the corresponding author.


\appendix 

\section{Details on the boxes used for calculating the flux densities}

\begin{figure}
\centering
\begin{tabular}{c}
\includegraphics[width=\columnwidth]{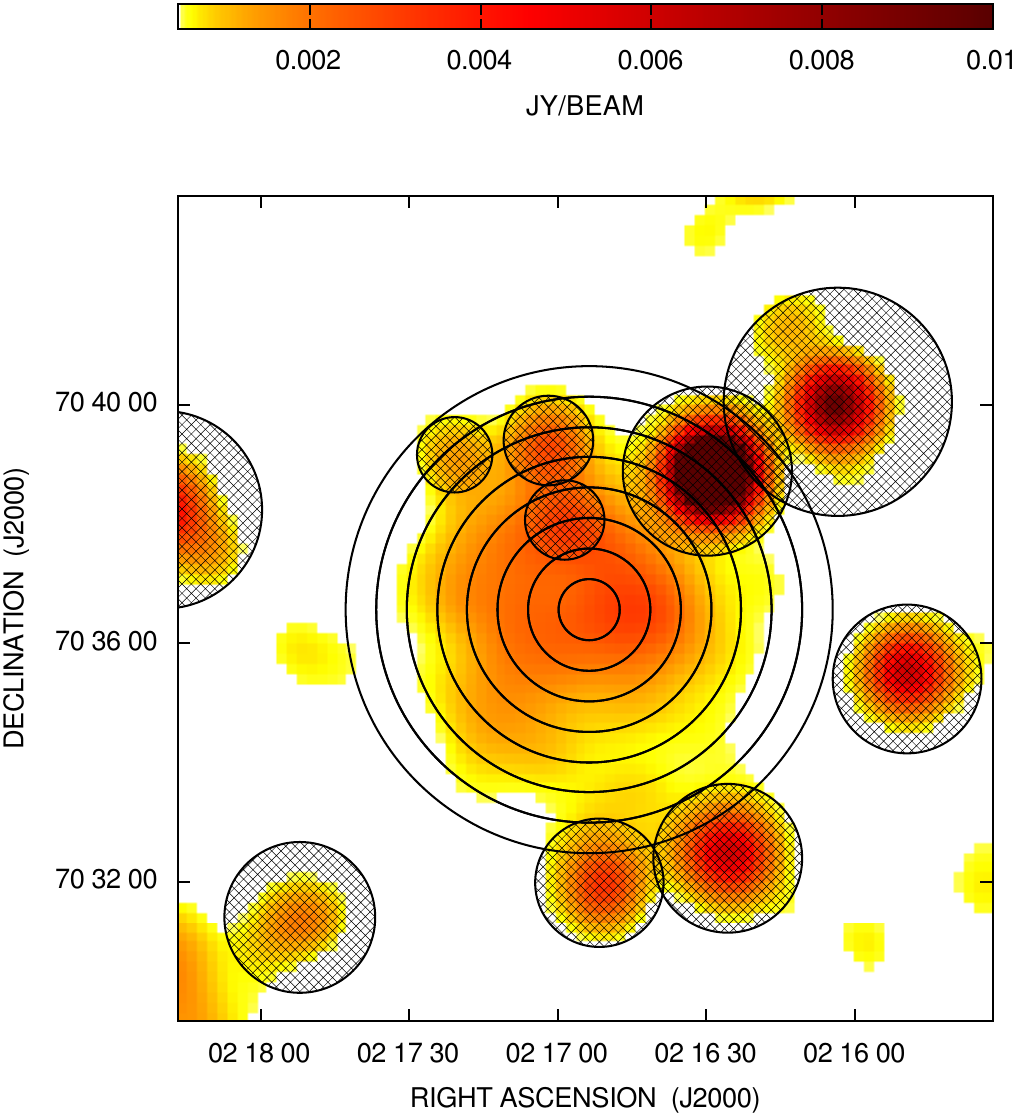}
\end{tabular}
\caption{Region of the radio halo in the combined SRT+VLA image with only pixels having surface brightness larger than $3\sigma$ (with $\sigma=0.15$ mJy/beam), where the concentric annuli used to calculate the surface brightness profile are shown. We also show the position and the extension of the regions we have blanked in order to subtract the contribution of the discrete sources.}
\label{fig.exp_masks}
\end{figure}

\begin{figure}
\centering
\begin{tabular}{c}
\includegraphics[width=\columnwidth]{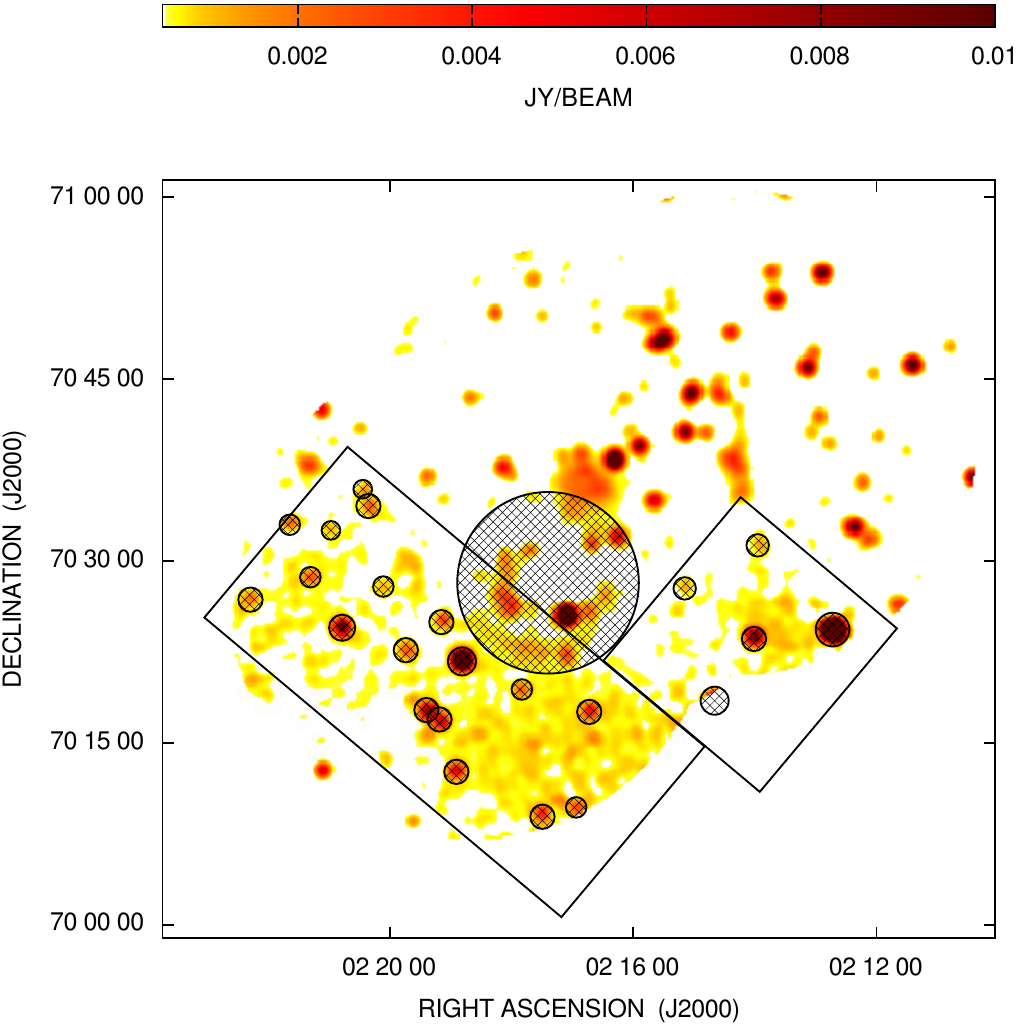}
\end{tabular}
\caption{South-eastern extended emission visible in the combined SRT+VLA image, where only pixels having surface brightness larger than $3\sigma$ (with $\sigma=0.15$ mJy/beam) are shown. Two rectangular regions within which we integrated the surface brightness to obtain the total flux are indicated. We also show the position and the extension of the regions we have blanked in order to subtract the contribution of the discrete sources and the cluster south-eastern relic.}
\label{fig.se_masks}
\end{figure}

In this Appendix we show the figures containing the boxes and the masks we have used to derive the values of the flux density for the regions we have discussed in the paper.
In particular, we show in Fig.\,\ref{fig.exp_masks} the concentric annuli used for performing the exponential fit of the radio halo (Section 4.1.1) and in Fig.\,\ref{fig.se_masks} the regions considered for deriving the flux density of the external south-eastern region (Section 4.2).

\bsp

\label{lastpage}

\end{document}